\documentclass[12pt,a4paper,dvips]{article}%
\usepackage{cite}
\usepackage[dvips]{graphicx}
\usepackage{palatino}
\usepackage{anysize}
\usepackage{fancyhdr}
\usepackage{amsmath}
\usepackage{amsfonts}
\usepackage{amssymb}
\usepackage[font=small]{caption}
\makeatletter \renewcommand\@biblabel[1]{} \makeatother

\marginsize{2.5 cm}{2.5 cm}{2.0 cm}{2.0 cm}
\RequirePackage{ifpdf}
\ifpdf
\RequirePackage[pdfpagelabels,pdfstartview=FitH,plainpages=false]{hyperref}
\pdfpagewidth=\paperwidth
\pdfpageheight=\paperheight
\else
\RequirePackage[dvips,pdfpagelabels,pdfstartview=FitH,plainpages=false]{hyperref}
\hypersetup{linktocpage}

\fi
\begin{document}

\title{Economic interactions and the distribution of wealth}
\author{Davide Fiaschi - Matteo Marsili\thanks{Respectively, Dipartimento di Scienze
Economiche, University of Pisa, Via Ridolfi 10, 56124 Pisa Italy, e-mail:
dfiaschi@ec.unipi.it and The Abdus Salam International Centre for Theoretical
Physics, Strada Costiera 11, 34014 Trieste Italy, email:
marsili@ictp.trieste.it. We thank the seminars' participants in Bologna, Pisa and Trento, and Anthony Atkinson and Vincenzo Denicol\`{o}  for useful comments. Usual disclaimers apply.}}
\maketitle

\begin{abstract}
This paper analyzes the equilibrium distribution of wealth in an economy where firms' productivities are subject to idiosyncratic shocks, returns on factors are determined in competitive markets, dynasties have linear consumption functions and government imposes taxes on capital and labour incomes and equally redistributes the collected resources to dynasties. The equilibrium distribution of wealth is explicitly calculated and its shape crucially depends on market incompleteness. In particular, a Paretian law in the top tail only arises if capital markets are incomplete. The Pareto exponent depends on the saving rate, on the net return on capital, on the growth rate of population and on portfolio diversification. On the contrary, the characteristics of the labour market mostly  affects the bottom tail of the distribution of wealth. The analysis also suggests a positive relationship between growth and wealth inequality.
\end{abstract}

\noindent {\em Keywords}: wealth distribution, incomplete markets, earnings distribution, capital income taxation, productivity shocks, portfolio diversification

\noindent {\em JEL classification numbers}: D1, D31, D33, N30

\newpage

\section{Introduction}

The statistical regularities in the distribution of wealth have attracted considerable interest since the pioneering works of \cite{Pareto1987} (see \cite{AtkinsonHarrison78}
and \cite{DaviesShorrocks1999} for a review). The efforts of economists have focused primarily on the understanding the micro-economic causes of inequality. A more recent trend, reviewed in \cite{ChatterjeeEtAl2005}, has instead focused on mechanistic models of wealth exchange with the aim of reproducing the observed empirical distribution. A general conclusion is that the Pareto distribution arises from the combination of a multiplicative accumulation process, and an additive term. 

This paper attempts to establish a link between these two literatures, by showing 
that the same mathematical structure emerges in a model which takes into account explicitely the complexity of market interactions of a large economy. In brief, the model describes how 
idiosyncratic shocks in the production of firms propagating through the financial and the labor markets shape the distribution of wealth. Market networks, i.e. who works and who invests in each firm, play a crucial role in determining the outcome. 
As suggested in \cite{Aiyagari1994}, the shape of the equilibrium distribution crucially depends on market incompleteness, i.e. on the fact that individuals do not invest in all firms. With complete markets, 
the equilibrium distribution of wealth is determined solely by shocks transmitted through the labor market, and it takes a Gaussian shape, a result at odds with empirical evidence (see, e.g., \cite{KlassEtAl(2006)}). Only when frictions and transaction costs impede full diversification of dynasties' portfolios, the shape of the top tail of the distribution follows a Paretian law. 
The Pareto exponent computed explicitly allows to individuate the effects which different parameters have on wealth inequality. We find that an increase in the taxation of capital income or in the diversification of dynasties' portfolios increases the Pareto exponent, whereas changes in the saving rate or in the growth rate of the population impact inequality in different ways, depending on technological parameters, due to {\em indirect} effects on the return on capital.

The bottom tail of the equilibrium distribution of wealth is instead crucially affected by the characteristics of labour market. With a labour market completely decentralized, so that individual wages immediately respond to idiosyncratic shocks to firms, the support of the equilibrium distribution of wealth includes negative values; on the contrary if all workers receive the same
wage, i.e. bargaining in the labour market is completely centralized, shocks are only transmitted through return on capital and the distribution of wealth is bounded away from zero.

Finally, we show that, if the growth rate of the economy is endogenous, there is a negative relationship between the latter and the Pareto exponent, i.e. wealth inequality.

\section{The Model}

We model a competitive economy in which $F$ firms demand capital and
labour. We assume all the wealth is owned by $N$ households (assumed to be infinitely lived), who 
offer capital and labour and decide which amount of their disposable income is
saved. Wages and returns on capital adjust to clear the labour and
capital markets respectively. 

We derive continuum time stochastic equations for the evolution of the distribution of wealth,
specifying the dynamics over a time interval $[t,t+dt)$ and then letting $dt\rightarrow0$. We refer the interested reader to \cite{FiaschiMarsili09} for details, and report directly the dynamical equations.
The wealth $p_i$ of household $i$ obeys the following stochastic differential
equation:
\begin{equation}
\frac{dp_{i}}{dt}=s\left[  \left(  1-\tau_{k}\right)  \rho p_{i}+\left(
1-\tau_{l}\right)  \omega l_{i}+\tau_{k}\rho\bar{p}+\tau_{l}\omega\bar
{l}\right]  -\chi-\nu p_{i}+\eta_{i}, \label{evolutionWealthAgent_i}%
\end{equation}
where $\eta_{i}$ is a white noise term with $E\left[  \eta_{i}\left(
t\right)  \right]  =0$ and covariance:%
\begin{equation}
E\left[  \eta_{i}\left(  t\right)  \eta_{i^{\prime}}\left(  t^{\prime}\right)
\right]  =\delta\left(  t-t^{\prime}\right)  H_{i,i^{\prime}}\left[  \vec
{p}\right]  , \label{covarianceMatrixEvolutionWealthAgent_i}%
\end{equation}
The first three terms in the r.h.d. of Eq. (\ref{evolutionWealthAgent_i}) detail a simple behavioral model of how the consumption of household $i$ depends on her income and wealth. The term in square brackets represents the disposable income of household $i$, which arises {\em i)} from the return on investment, at an interest rate $\rho$, taxed by government at a flat rate $\tau_k$, and {\em ii)} from income from labor, which is taxed at a rate $\tau_l$. Here $\omega$ is the wage rate and $l_i$ is the labor endowment of household $i$. The last two terms in the square brackets denote the equal redistribution of collected taxes on capital and labor markets, respectively, where $\bar p$ and $\bar l$ are the average wealth and labor endowment. A fraction $s$ of the income is saved, i.e. $s$ is the saving rate on income.
The term $\chi$ represents minimal consumption, i.e. the rate at which household would consume in the absence of wealth and income, whereas $\nu$ is the rate of consumption of wealth. 
This simple consumption model finds solid empirical support, as discussed in  \cite{FiaschiMarsili09}.

The return of capital markets $\rho$ and the wage rate $\omega$ are fixed by the equilibrium conditions of the economy. In brief, each firm $j$ buys capital $k_j$ and labor $l_j$ from households in capital and labor markets, i.e.:
\[
k_j=\sum_{i=1}^N \theta_{i,j} p_i,~~~~l_j=\sum_{i=1}^N\phi_{i,j},~~~~j=1,\ldots,F,
\]
where $\theta_{i,j}$ ($\phi_{i,j}$) is the fraction of $i$'s wealth (labor) invested in firm $j$. These are used as inputs in the production of firm $j$, and produce an  amount $dy_j=q(k_j,l_j)dA_j$ of output in the time interval $dt$. Here $q(k,l)$ is the production function of firms, whereas $dA_j(t)$ is an idiosyncratic shock,  which is modeled as a random variable with mean $E[dA_j]=adt$ and variance $a^2\Delta dt$.

Under the standard assumption that $q(k,l)=lg(k/l)$ is an homogeneous function of degree 
one, when capital and labor markets clear, we find that {\em i)} each firm has the same capital to labor ration $k_j/l_j=\lambda$, {\em ii)} the return on capital is given by $\rho=ag'(\lambda)$ and {\em iii)} the wage rate is $\omega=a[g(\lambda)-\lambda g'(\lambda)]$. Since labor and capital are provided by households, and because of {\em i)}, the constant $\lambda=\bar p$ also equals household wealth per unit labor. Setting $l_i=1$ for all $i$, the constant $\lambda$ then equals the average wealth $\bar p$ of households.

The covariance of the stochastic noise in Eq. (\ref{evolutionWealthAgent_i}) is given by:
\begin{align*}
H_{i,i^{\prime}}\left[  \vec{p}\right]   &  =\Delta s^{2}\left\{  (1-\tau
_{k})^{2}\rho^{2}p_{i}p_{i^{\prime}}\Theta_{i,i^{\prime}}+\left(  1-\tau
_{l}\right)  ^{2}\omega^{2}l_{i}l_{i^{\prime}}\Phi_{i,i^{\prime}}\right.  +\\
&  +(1-\tau_{k})(1-\tau_{l})\rho\omega\left[  p_{i}l_{i^{\prime}}%
\Omega_{i,i^{\prime}}+l_{i}p_{i^{\prime}}\Omega_{i^{\prime},i}\right]  +\\
&  +\frac{\tau_{k}\rho+\tau_{l}\omega/\lambda}{N}\left[  (1-\tau_{k}%
)\rho(p_{i}\vartheta_{i}+p_{i^{\prime}}\vartheta_{i^{\prime}})+(1-\tau
_{l})\omega(l_{i}\varphi_{i}+l_{i^{\prime}}\varphi_{i^{\prime}})\right]  +\\
&  \left.  +\frac{\left[  \tau_{k}\rho+\tau_{l}\omega/\lambda\right]  ^{2}%
}{N^{2}}\sum_{j=1}^{F}k_{j}^{2}\right\}  ,
\end{align*}
where
\begin{equation}
\vartheta_{i}=\sum_{i^{\prime}=1}^{N}\Theta_{i,i^{\prime}}p_{i^{\prime}%
},~~~\varphi_{i}=\sum_{i^{\prime}=1}^{N}\Omega_{i,i^{\prime}}p_{i^{\prime}}.
\label{parameterPropWealthDynamicsAgenti_bis}%
\end{equation}
and
\begin{equation}
\Theta_{i,i^{\prime}}=\sum_{j=1}^{F}\theta_{i,j}\theta_{i^{\prime}%
,j},~~~\Omega_{i,i^{\prime}}=\sum_{j=1}^{F}\theta_{i,j}\phi_{i^{\prime}%
,j}\ \text{and }\Phi_{i,i^{\prime}}=\sum_{j=1}^{F}\phi_{i,j}\phi_{i^{\prime
},j}. \label{parametersPropWealthDynamicisAgenti}%
\end{equation}
The parameters in Eq. (\ref{parameterPropWealthDynamicsAgenti_bis}) characterize the degree of intertwinement of economic interactions, i.e.
how random shocks {\em propagate} throughout the economy. For example
$\Theta_{i,i^{\prime}}$ is a scalar which represents the overlap of
investments of dynasty $i$ with those of dynasty $i^{\prime}$.

\section{Infinite Economy}

We analyze the properties of the stochastic evolution of wealth discussed in the previous paragraph in the case of an infinite economy, that is of
an economy where $N$ and $F\rightarrow\infty$. In particular, we assume that
$F=fN$, where $f$ is a positive constant. This assumption is not a relevant
limitation of the analysis because in a real economy $N$ and $F$ may be of the
order of some millions. We take the further simplifying assumption that households do not differ among
themselves in their endowment of labour $l_{i}$, in the diversification of their portfolios $\Theta_{i,i}$, in
the allocation of their wealth among the firms where they are working
$\Omega_{i,i} $ and in the number of firms where they are working $\Phi_{i,i} $, i.e. we assume that: $l_{i}   =\bar{l}=1,~ \Theta_{i,i}   =\bar{\Theta},~\Omega_{i,i}   =\bar{\Omega}$ and $\Phi_{i,i}   =\bar{\Phi}\quad\forall i$.
For example, $\bar\Theta=1$ implies no diversification of the dynasties' portfolios (i.e. all wealth is invested in the same firm), whereas $\Theta=1/F$ (i.e. $\Theta \to 0$ for $F \to \infty$) corresponds to maximal diversification of portfolios; similarly, $\Phi=1$ means that each dynasty is working in just one firm.

In the limit $N,F\to\infty$, the per capita wealth $\bar{p}$ follows a deterministic dynamics given
by%
\begin{equation}
\frac{d\bar{p}}{dt}=s\left(  \rho\bar{p}+\omega\right)  -\chi-v\bar{p}.
\label{averageWealthDynamicsLargeN}%
\end{equation}
Besides a technical condition\footnote{The technical condition $\sum_{i=1}^{N}\theta_{i,j}\leq\bar{\theta} \; \forall j,N$ is needed to show this result.}, this result requires that the average wealth satisfies the Law of Large Numbers, i.e. that the wealth distribution $f(p)$ has a finite first moment.

Two different regimes are possible: {\em i)} the stationary economy where wealth is constant in equilibrium; and {\em ii)} the endogenous growth economy, where wealth is growing at constant rate in equilibrium.

\subsection{Stationary Economy}

If the growth rate of per capita wealth becomes negative for large
value of $\bar{p}$, i.e. if 
\begin{equation}
\lim_{\bar{p}\rightarrow\infty}\text{ }g^{\prime}\left(  \bar{p}\right)
<\frac{\nu}{sa}, \label{existence cond2}%
\end{equation}
then the economy approaches a stationary state.\footnote{For the proof see \cite{FiaschiMarsili09}.} In this case, the distribution of wealth depends on the parameters $\bar\Theta,\bar\Phi$ and $\bar\Omega$:

\begin{itemize}

\item In an infinite economy when household can fully diversify both their income from capital investment and labour (i.e. $\theta_{i,j}=\phi_{i,j}=1/F$), they can eliminate all sources of risk, i.e. $\bar{\Theta}, \bar{\Omega}= \bar{\Phi}=0$. Therefore their income is deterministic and, in  equilibrium, they all end up with the same wealth, i.e. $p_i=\bar{p}$. Therefore, if $\bar{\Theta}, \bar{\Omega}= \bar{\Phi}=0$ (complete markets) then:
\begin{equation}
f\left(  p_{i}\right)  = \delta\left(p_i - \bar{p} \right).
\label{equilibriumDistributionp_iCompleteMarkets}
\end{equation}

\item When households can fully diversify their portfolios ($\theta_{i,j}=1/F$), but they work in a limited number of firms, the wealth distribution is determined by the uninsurable idiosyncratic shocks
arising from labour income. In this case, in the infinite economy, $\bar{\Theta}, \bar{\Omega}=0$ and $\bar{\Phi} >0$ and, the equilibrium distribution of wealth attains a Gaussian shape,
\begin{equation}
f\left(  p_{i}\right)  = \mathcal{N} e^{-\frac{\left(z_{0}-z_{1}p_{i}\right)^2}{z_1 a_0}},
\label{equilibriumDistributionp_iCompleteFinancialMarkets}
\end{equation}
with mean $z_0/z_1=\bar{p}$ and variance $a_0/\left(2 z_1\right)$ (these parameters are defined below in Eq. (\ref{equilibriumDistributionp_i})).

\item In the more realistic incomplete market case, i.e. $\bar{\Theta}, \bar{\Omega}, \bar{\Phi} >0$, i.e. when full diversification is not possible, both in the capital and in the labor market (incomplete markets), then:
\begin{equation}
f\left(  p_{i}\right)  =\left[  \frac{\mathcal{N}}{\left(  a_{0}+a_{1}%
p_{i}+a_{2}p_{i}^{2}\right)  ^{1+z_{1}/a_{2}}}\right]  e^{4\left[  \frac
{z_{0}+z_{1}a_{1}/\left(  2a_{2}\right)  }{\sqrt{4a_{0}a_{2}-a_{1}^{2}}%
}\right]  \arctan\left(  \frac{a_{1}+2a_{2}p_{i}}{\sqrt{4a_{0}a_{2}-a_{1}^{2}%
}}\right)  }; \label{equilibriumDistributionp_i}%
\end{equation}
where%
\begin{align*}
z_{0}  &  =s\left[  \omega^{\ast}+\tau_{k}\rho^{\ast}\bar{p}\right]  -\chi;\\
z_{1}  &  =\nu-s\left(  1-\tau_{k}\right)  \rho^{\ast};\\
a_{0}  &  =\Delta s^{2}\left(  1-\tau_{l}\right)  ^{2}\omega^{\ast2}\bar{\Phi
};\\
a_{1}  &  =2\Delta s^{2}(1-\tau_{k})(1-\tau_{l})\rho^{\ast}\omega^{\ast}%
\bar{\Omega} \; \text{ and}\\
a_{2}  &  =\Delta s^{2}\left(  1-\tau_{k}\right)  ^{2}\rho^{\ast2}\bar{\Theta},%
\end{align*}
where $\mathcal{N}$ is a constant defined by the condition $\int_{-\infty
}^{\infty}f\left(  p_{i}\right)  dp_{i}=1$. For large $p_{i}$ $f\left(  p_{i}\right) \sim p_i^{-\alpha-1} $ follows a Pareto distribution
whose exponent is given by:%
\begin{equation}
\alpha=1+2z_{1}/a_{2}=1+2\frac{\nu-s\left(  1-\tau_{k}\right)  \rho^{\ast}%
}{\Delta s^{2}(1-\tau_{k})^{2}\rho^{\ast}{}^{2}\bar{\Theta}}.
\label{ParetoParameterStationaryEconomy}%
\end{equation}
We observe that $z_{1}$, $a_{2}>0$ (see Eq. \ref{existence cond2})  and hence $\alpha>1$: this ensures that the first moment of the wealth distribution is indeed finite. 

\item The case $\bar \Theta>0$ and $\bar\Phi=\bar\Omega=0$ corresponds to the rather unrealistic situation where households distribute their labor on all firms. It turns out, however, that the resulting distribution of wealth is exactly the same as that of an economy in which Trade Unions have a very strong market power, such that the bargaining on labour market is completely centralized. Hence wages are fixed ({\em staggered wages}) in the short run and productivity shocks are absorbed by the returns on capital. Mathematically this corresponds exactly to the case $\bar\Phi=\bar\Omega=0$, for which the distribution of wealth reads 
\begin{equation}
f\left(  p_{i}\right)  =\frac{\mathcal{N}}{a_{2}p_{i}^{2\left(
1+z_{1}/a_{2}\right)  }}e^{-\left(  \frac{2z_{0}}{a_{2}p_{i}}\right)
},\label{equilibriumDistributionp_iStaggeredWages}%
\end{equation}
where $\mathcal{N}$ is a normalization constant, $z_{1}$ and $a_{2}$ are the same as above.

\end{itemize}

The results above indicate that while the bottom of the wealth distribution is determined by the labor market, the top tail only depends on the working of capital markets. If wages respond to productivity shocks and households are not able to fully diversify their employment (as is typically the case), then the distribution extends to negative values of the wealth. If, instead, staggered wages are imposed by a centralized bargaining in the labor market, then inequality in the bottom tail is highly reduced.

With respect to the upper tail, we
observe that the assumption  $\Theta_{i,i}   =\bar{\Theta} \quad\forall i$ eliminates cross-household
heterogeneity in Eq. (\ref{equilibriumDistributionp_i}). However, it is worth noting that if dynasties were heterogeneous in their portfolio diversification, i.e. $\Theta_{i,i} \neq \Theta_{i',i'}$, then the top tail distribution would be populated by the dynasties with the highest $\Theta_{i,i}$, that is by those dynasties with the less diversified portfolios. This finding agrees with the empirical evidence on the low diversification of the portfolios of wealthy households discussed in \cite[Cap. 10]{GuisoEtAl2001}.

The (inverse of the) exponent $\alpha$ provides a measure of inequality. Our results show that inequality increases with the volatility $\Delta$ of productivity shocks and with the concentration $\bar{\Theta}$ of household portfolios, and it decreases with capital taxation
$\tau_{k}$. 

Changes in $s$ and $v$ have, on the contrary, an ambiguous effect on the size of the top tail of distribution of wealth. More precisely, an increase in the gross return on capital
$\rho^{\ast}$ amplifies inequality (i.e. $\partial\alpha/\partial
\rho^{\ast}<0$). When $s$ increases a {\em direct} effect tends to decrease $\alpha$, while an {\em induced} effect tends to increase $\alpha$, because it causes an increase in the
equilibrium per capita wealth $\bar{p}^{\ast}$, and hence a decrease in the
return on capital $\rho^{\ast}$. When $\nu$ increases the
contrary happens. Without specifying the technology $g(\lambda)$ it is not possible to
determine which effect prevails (see \cite{FiaschiMarsili09} for some examples).

\subsection{Endogenous Growth Economy\label{secEndoGrowthRate}}

If the dynamics of per capita wealth obeys Eq. (\ref{averageWealthDynamicsLargeN}) and
\begin{equation}
\lim_{\bar{p}\rightarrow\infty}g^{\prime}\left(  \bar{p}\right)  >\frac{v}%
{sa},\label{condPropExpandingEconomy}%
\end{equation}
then, in the long run, the returns on factors are given by:%
\begin{align}
\rho^{\ast}=\lim_{\bar{p}\rightarrow\infty}ag^{\prime
}\left(  \bar{p}\right)  \text{ and}\label{interestRateExpandingEconomy}\\
\omega^{\ast}=0.\label{wageExpandingEconomy}%
\end{align}
and per capita wealth grows at the rate\footnote{If $g\left(  0\right)  > \chi/ \left( sa \right)$, this result holds independently of the initial level of per capita wealth, otherwise endogenous growth sets in only if the initial per capita wealth is sufficient high (see \cite{FiaschiMarsili09}).}
\begin{equation}
\psi^{EG}=\lim_{\bar{p}\rightarrow\infty}sag^{\prime}\left(  \bar{p}\right) -\nu=s\rho^{\ast}-\nu.
\label{growthRateLongRun}%
\end{equation}

Notice that $\psi^{EG}$ is independent of the flat tax rate on capital\footnote{This is due to the assumption of constant saving rate $s$. Generally, $s$ increases with the net return on capital $\left(  1-\tau_{k}\right) \rho^{\ast}$, hence $s$ decreases with $\tau_{k}$. This suggests that the growth rate $\psi^{EG}$ decreases with capital taxation $\tau_k$.} $\tau_{k}$ and of the diversification of dynasty $i$'s
portfolio $\bar{\Theta}$; however, $\psi^{EG}$ increases with saving rate $s$ and with return on capital $\rho^{\ast}$ and it decreases with $\nu$; changes in technology which increase the return on capital, therefore, also cause an increase in $\psi^{EG}$. 

The distribution of wealth is best described in terms of 
the relative per capita
wealth of households $u_{i}=p_{i}/\bar{p}$. In the long run household $i$'s relative
wealth obeys the following stochastic differential equation:%
\begin{equation}
\lim_{t\rightarrow\infty}\frac{du_{i}}{dt}=s\rho^{\ast}\tau_{k}%
(1-u_{i})+\tilde{\eta}_{i}, \label{accEqEndoGrowth}%
\end{equation}
where $\tilde{\eta}_{i}=\eta_{i}/\bar{p}$ is a white noise term with $E\left[
\tilde{\eta}_{i}\left(  t\right)  \right]  =0$ and covariance:%
\begin{equation}
E\left[  \tilde{\eta}_{i}\left(  t\right)  \tilde{\eta}_{i^{\prime}}\left(
t^{\prime}\right)  \right]  =\delta\left(  t-t^{\prime}\right)  H_{i,i^{\prime
}}\left[  \vec{u}\right]  , \label{covarianceEndoGrpwth}%
\end{equation}
where:%
\[
\lim_{t\rightarrow\infty}\lim_{N\rightarrow\infty}H_{i,i^{\prime}}%
[\vec{u}]=\left[  \Delta s^{2}(1-\tau_{k})^{2}\rho^{\ast}{}^{2}\Theta
_{i,i^{\prime}}\right]  u_{i}u_{i^{\prime}}.
\]

In the limit $\bar{p}\rightarrow\infty$ the equilibrium wage rate converges to $0$ and therefore wages do not play any role in the dynamics of relative per capita wealth of dynasty $i$, as stated above. In the long run, the equilibrium distribution of the relative per capita wealth $u_{i}$, in the non-trivial (and realistic) case of incomplete markets $\bar{\Theta}>0$, is given by
\begin{equation}
f^{EG}(u_{i})=\frac{\mathcal{N}^{EG}}{u_{i}^{\alpha^{EG}+1}}e^{-(\alpha
^{EG}-1)/u_{i}}, \label{marginalDistributionExpandigEconomy}%
\end{equation}
\bigskip where $\mathcal{N}^{EG}$ is a normalization constant, and%
\begin{equation}
\alpha^{EG}=1+2\frac{\tau_{k}}{\Delta s(1-\tau_{k})^{2}\rho^{\ast}\bar{\Theta
}} \label{alphaExpandingEconomy}%
\end{equation}
is the Pareto exponent.

We remark that while capital taxation $\tau_k$ has no direct effect on growth, it has a direct effect on inequality.\footnote{The results above, in the limit $\tau_{k}\rightarrow0$, do not reproduce the behavior  of the economy with $\tau_{k}=0$:  Indeed, Eq. (\ref{accEqEndoGrowth}), with $\tau_{k}=0$ and $H_{i,i'}=0$ for $i \not= i'$, describes independent log-normal processes $u_i\left(t\right)$.} Hence capital taxes do not (directly) affect growth, but have a crucial redistributive function: wealth is redistributed away from wealthy to poor dynasties by an amount proportional to aggregate wealth, so preventing the possible ever-spreading wealth levels, and stabilizing the equilibrium distribution of relative wealth.

Finally, the Pareto exponent is continuous across the transition from a stationary to an endogenously growing economy, i.e.
\[
\lim_{s\rho^{\ast}-\nu\rightarrow0^{-}}\alpha=\lim_{s\rho^{\ast}%
-\nu\rightarrow0^{+}}\alpha^{EG},
\]
though it has a singular behaviour in the first derivative (with respect to
$\nu$ or $s$). We remark that the Pareto exponent $\alpha^{EG}$ decreases with
saving rate $s$, return on capital $\rho^{\ast}$, the diversification of portfolio $\bar{\Theta}$ and it increases with $\tau_{k}$; $\alpha^{EG}$ is, on the contrary, independent of $\nu$.

\begin{figure}[hptb]
\centering
\includegraphics[width=7cm]{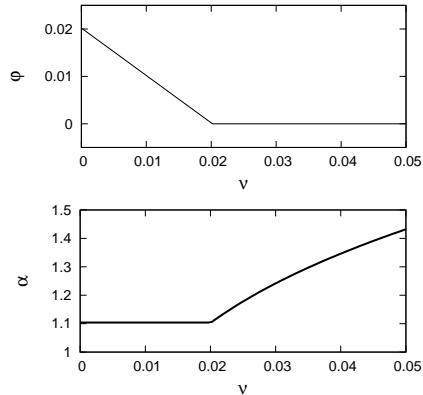}
\caption{Behavior of the Pareto exponent as a function of the parameter $\nu$ for an economy where $g(\lambda)=[\epsilon\lambda^\gamma+1-\epsilon]^{1/\gamma}$ (constant elasticity of substitution technology) with $\epsilon=0.2$ and $\gamma=0.7$. The other parameters take values: $a=1.0, 
s=0.2, \tau_k=0.2$ and $\Delta\Theta=300$.}
\label{fig}
\end{figure}

Interestingly, since $\psi^{EG}$ increases with $s$
and $\rho^{\ast}$, we find an inverse
relationship between growth and wealth inequality. Indeed the Pareto exponent $\alpha^{EG}$ and the growth rate $\psi^{EG}$ show an inverse relationship under changes in saving rate $s$ and/or return on capital $\rho^{\ast} $. For example, an economy increasing its saving rate $s$ (or its return on capital $\rho^{\ast}$) should move to an equilibrium where both its growth rate and its wealth inequality (in the top tail of the distribution of wealth) are larger than before. The behavior of the Pareto exponent and of the growth rate is illustrated in Fig. \ref{fig} for a particular choice of the production function.

\section{Conclusions and future research}

This paper discusses how the equilibrium distribution of
wealth can be derived from the equilibrium of an economy with a large number of firms and households, who interact through the capital and the labour markets. Under incomplete markets, the top tail of the equilibrium distribution of wealth is well-represented by a Pareto distribution, whose exponent depends on the saving rate, on the net return on capital, on the growth rate of the population, on the tax on capital income and on the degree of diversification of portfolios. 
On the other hand, the bottom tail of the distribution mostly depends on the working of the labour market: a labour market with a centralized bargaining where workers do not bear any risk determines a lower wealth inequality.

Our framework neglects important factors which have been shown to
have a relevant impact on the distribution of wealth (see 
\cite{DaviesShorrocks1999}). Moreover, our analysis is relative to the
equilibrium distribution of wealth and it neglects out-of-equilibrium
behaviour and issues related to the speed of convergence.  The relationship between the distribution of wealth and the distribution of income, as well as its relation with the distribution of firm sizes is a further interesting extension of our analysis.

An additional interesting aspect is that of finite size effects in aggregate fluctuations. This issue has been recently addressed by \cite{Gabaix08} in an economy in which aggregate wealth exhibits a stochastic behaviour. In the light of our findings, the latter behaviour can arises because of correlations in productivity shocks, which were neglected here, because dynasties concentrate their investments in few firms/assets or
because the number of firms/assets is much smaller than the number of dynasties. This extension would draw a theoretical link between the dynamics of the distribution of wealth, the distribution of firm size and business cycle.


\begin{thebibliography}{9999999999999999999999999999999999999999999999}                                                   %


\bibitem[Atkinson and Harrison (1978)]{AtkinsonHarrison78}Atkinson A. B., and
A.J. Harrison (1978), Distribution of the Personal Wealth in Britain, Cap. 3,
Cambridge: Cambridge University Press.



\bibitem[Aiyagari (1994)]{Aiyagari1994} Aiyagari R. (1994), Uninsured Idiosyncratic Risk and Aggregate Saving, Quarterly Journal of Economics, 109, 659-684.













\bibitem[Chatterjee et al. (2005)]{ChatterjeeEtAl2005}Chatterjee A., S.
Yarlagadda and B.K. Chakrabarti (eds) (2005) Econophysics of Wealth
Distribution, Berlin: Springer.




\bibitem[Davies and Shorrocks (1999)]{DaviesShorrocks1999}Davies, J.B. and .
F. Shorrocks (1999), The Distribution of Wealth in Handbook of Income
Distribution, A.B. Atkinson and F. Bourguignon (eds), Amsterdam: Elsevier.





\bibitem[Gabaix (2008)]{Gabaix08} Gabaix X. (2008), TheGranularOriginsofAggregateFluctuations, 
SSRN working paper: http://ssrn.com/abstract=1111765.

\bibitem[Fiaschi and Marsili (2009)]{FiaschiMarsili09} Fiaschi D and Marsili M, (2009), Distribution of Wealth and Incomplete Markets: Theory and Empirical Evidence, DSE Discussion Paper 2009/83, University of Pisa, Italy, (available at http://ideas.repec.org/p/pie/dsedps/2009-83.html).




\bibitem[Guiso et al. (2001)]{GuisoEtAl2001}Guiso L., M. Haliassos and T. Jappelli (eds) (2001). Household Portfolios, Cambridge: MIT Press.







\bibitem[Klass et al. (2006)]{KlassEtAl(2006)}Klass O., Biham O., Levy M., O.
Malcai and S. Solomon, The Forbes 400 and the Pareto wealth distribution,
Economics Letters, 90, 290-295.






\bibitem[Pareto (1897)]{Pareto1987}Pareto, V. (1897), Corso di Economia
Politica, Busino G., Palomba G., edn (1988), Torino: UTET.
















\end{thebibliography}
\end{document}